# Optimized Active Cooling and Refrigeration using Antidoted Graphene for Heat Management of Microelectronics

Shuang Tang[1,*], Andy Juan[1], David Drysdale[1], Joseph Duarte Menjivar[1], Jason Guzman[1]

[1] *1College of Engineering, State University of New York Polytechnic Institute, Albany/Utica, NY, 12203/13502, USA*

*** E-mail**: *tangs1@sunypoly.edu*

## Abstract

With the technology of artificial defects creating, we can tune the band structure and transport properties of many two-dimensional (2D) layered materials. One prototype materials system is the antidoted graphene sheet, where periodical pores are made using focuses ion or electron beams in the nanoscale. We here study the electrical conductivity, thermopower, and active rates of cooling and refrigeration of antidoted graphene samples with different pore-radii and interporous distances. We use a calculation method that takes into consider the sensitivity of transport to charge carrier energy, which can be used to describe the elastic and inelastic scatterings in diffusive, ballistic and quantum hopping regimes. It is found that our results from the new calculational approach are more consistent with the experimental data, compared to some traditional methodologies. It is also interesting to see that the optimized active rates of cooling and refrigeration are very robust against the distribution variations of interporous distance and the pore-radius, which implies easy industrialization and inexpensive manufacturing. The same analysis and investigation can also be extended to many other layered materials, including the transitional metal dichalcogenides (TMD), blue phosphorene, and tellurium.

**1. Introduction**

As we are reaching the limit of Moore's law [1] in upgrading computer chips based on the silicon materials system, many new candidates have been proposed for the next generation microelectronics, including graphene [2], blue phosphorene [3], tellurene [4], and other low-dimensional materials [5, 6]. Besides the power consumption and computational frequency, the heat management of such microsystems is also an important issue that needs to be resolved. Beyond the micro-fluid systems for passive heat management, researchers have been interested in solid state thermoelectric cooling and refrigeration technology [7], which is non-noisy, non-fluid, compact, and easy to cascade. Many materials have been suggested to make thermoelectric cooling/refrigeration systems, including bismuth antimony thin films [8-10], silicon germanium nanowires [11, 12], lead telluride nanocomposites [13, 14], and others [15], among which the graphene-based derivatives are attracting intensive interests [16-18], which can be easily compacted with other layered materials of novel electronics.

Antidoted graphene [19, 20] are nanoporous materials that can be used to tune the band structure, the scattering mechanism and thus, the transport properties of pristine graphene. Nanoscale pores can be made on a graphene layer using extreme ultraviolet light [21, 22], focused ion beam [23, 24], or electron beam lithography [25, 26], periodically or randomly. This subtractive process of pores introducing is referred to as antidoting. Xu. et al. [26] has realized a thermopower of 26.3 μW/cm·K$^2$ at room temperature and 3.01 μW/cm·K$^2$ at the cryogenic temperature of 82 K, before optimization. Unfortunately, there is not a more detailed study on the optimized performance in active cooling and refrigeration using antidoted graphene sheet at both the room and cryogenic temperatures.

The calculation of active cooling and refrigeration rates largely depends on the electrical conductivity and the thermopower. Many methodologies have been proposed and utilized in the calculation of electrical conductivity and thermopower for graphene and other materials systems, among which the semi-classical Boltzmann transport theory using the code of *BoltzTrap* code [27] and the maximally localized Wannier function method using the package of *Electron-Phonon Wannier* (*EPW*) [28] package are with special importance. The former is simplifying the calculation using a constant relaxation time approximation, and the latter is neglecting scatterings from mechanisms other than the electron-phonon interaction. However, as Tang et al. [26, 29, 30] pointed out, the optimal thermopower is highly depending on the specific carrier scattering mechanism(s), and hence the sensitivity of transport/scattering rate to the carrier energy. Therefore, a further advanced method should be employed to consider the influence of such energy sensitivity on the optimal thermopower, and thus, the active cooling and refrigeration rates.

In this paper, we will study the conductivity and thermopower by taking the carrier scattering mechanism induced energy sensitivity into consideration with comparison to the *BoltzTrap* method [27], the *EPW* method [28], and the experimental data [26]. The active cooling and refrigeration rates of antidoted graphene sheet with different pore-radii and interpore distances will be then studied for both the room and the cryogenic temperatures.

## 2. Method

The electrical conductivity and the thermopower can be generally calculated as [29],

$$\sigma = q^2 \sum_{Bands} \int q\theta \left(-\frac{\partial f}{\partial E}\right) dE, \quad (1)$$

and

$$S = -\frac{k_B}{q} \frac{\sum_{Bands} \int q\theta \left(-\frac{\partial f}{\partial E}\right)\left(\frac{E-E_f}{k_B T}\right) dE}{\sum_{Bands} \int q\theta \left(-\frac{\partial f}{\partial E}\right) dE}, \quad (2)$$

where $k_B$ is the Boltzmann constant, $q$ is the positive or negative elementary charge, $f$ is the probability distribution function of electrons and $E_f$ is the Fermi level. The transport rate $\theta$ measures the density of free charge carriers passing across a unit area per unit time, and can be determined using an iterative approach [31-36] to consider both elastic and inelastic scatterings in diffusive [37-40], ballistic [41-45] and quantum hopping transport regimes [46-49], as

$$f(\mathbf{k}) = f_0[\varepsilon(\mathbf{k})] + x\theta(\mathbf{k}), \quad (3)$$

where **k** is the lattice momentum, and $f_0$ is $f$ function when the system reaches an equilibrium, and $x$ is the angle between the momentum **k** and the externally exerted force. The perturbation $\theta(\mathbf{k})$ can be updated iteratively by:

$$\theta(\mathbf{k}) = \frac{\chi_{i,in}[\theta(\mathbf{k})]+\chi_{e,in}[\theta(\mathbf{k})]+\chi_{b,in}[\theta(\mathbf{k})]+\chi_{h,in}[\theta(\mathbf{k})]-v(\mathbf{k})\left(\frac{\partial f}{\partial z}\right)-\frac{e\varepsilon}{\hbar}\nabla_{\mathbf{k}}f}{\chi_{i,out}(\mathbf{k})+\chi_{e,out}(\mathbf{k})+\chi_{b,out}(\mathbf{k})+\chi_{h,out}(\mathbf{k})}, \qquad (4)$$

where the $\chi_{in,*}$ and $\chi_{out,*}$ terms are counting the probability increment caused by inward and outward transport, respectively. The subscripts of "$i$", "$e$", "$b$", and "$h$" stand for inelastic, elastic, ballistic and hopping scattering/transport, respectively. The term $\left(\frac{\partial f}{\partial z}\right)$ is counting the change of probability density caused by the temperature gradient along the z direction, and the term $\frac{e\varepsilon}{\hbar}\nabla_{\mathbf{k}}f$ is counting the effect of external electrical fields.

Tang et al. [26, 29, 30] has found that the energy sensitivity of carrier transport ($s$), which is representing the relative change of transport rate per unit relative change of carrier energy ($\varepsilon$) counting from the corresponding band edge $s=(d\theta/\theta)/(d\varepsilon/\varepsilon)$, is a key factor that ultimately determines the optimal value of the thermopower during the integration of Eq. (1) and (2), which is neglected in the *BoltzTrap* method [27] and the *EPW* [28] method. The value of such an energy sensitivity is determined by the specific materials system, the carrier scattering mechanisms, and the temperature.

Zebarjadi et al. [50] have pointed out that besides the passive thermal conductivity, the active cooling rate of a material caused by thermoelectric effect can be calculated by

$$r_{a,c} = \frac{\sigma S^2 T_H}{2\Delta T}, \qquad (5)$$

where $T_C$ and $T_H$ are the cold-side and hot-side temperatures, respectively, and $\Delta T = T_H - T_C$ is the temperature drop. The active refrigeration rate can, instead, be calculated by

$$r_{a,r} = \frac{\sigma S^2 T_c}{2\Delta T} \ . \tag{6}$$

The transport property of an antidoted graphene may be influenced by the shape of the pores, the pore-radius, and the interpore distance. Without loss of generality, we study the graphene samples antidoted with square pores, with different pore-radii and interpore distances as defined and illustrated in Figure 1.

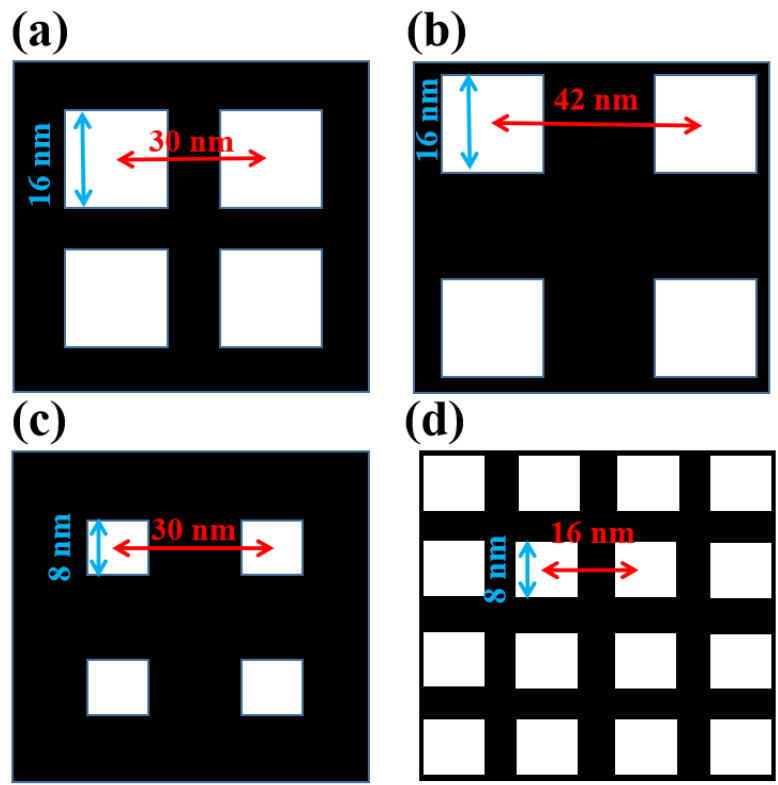

**FIGURE 1:** We are investigating antidoted graphene Samples 1-4 as illustrated in (a)-(d). The radii of the square pores are 8, 8, 4 and 4 nm, respectively, and the corresponding interporous distances are 30, 42, 30 and 16 nm.

### 3. Results and Discussions

From our previous study, we found that the energy sensitivity in squarely antidoted graphene is $s$=1.11 for the room temperature of 300 K and $s$=0.30 for the cryogenic temperature of 82 K [26]. The calculated thermopower of the antidoted graphene sample shown in Figure 1 (a) using our method is exhibited in Figure 2, compared to the other two traditional methods. The comparison is suggesting that although using the constant relaxation time approximation or using the phonon scattering approximation can capture the main trend of the thermopower in a large range, our method can better capture the details near the maximum-value range, due to the incorporation of the information on energy sensitivity. In this case, the actual energy sensitivity in the antidoted graphene sample is larger than ~0 as assumed in the *BoltzTrap* method and ~0.5 as assumed in the *EPW* method, which results in their underestimation of the optimal thermopower.

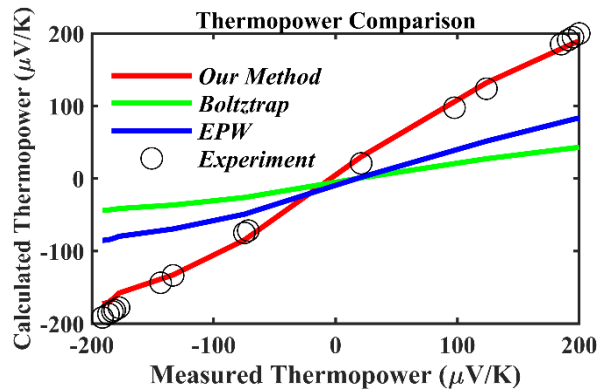

**FIGURE 2:** Comparison of thermopower of Sample 1 at the room temperature of $T$=300 K, calculated by our method, the *BoltzTrap* method, and the *EPW* method, with respect to the data measure from experiment by Xu et al. [26]

We then use our method to calculate the values of thermopower and power factor that are optimized against the gate voltage or Fermi level, for the four different antidoting cases shown in Figure 1. As we can see from Figure 3, the optimal values of thermopower differ noticeably in the four different samples. Since we can view the antidoted graphene as a network of nanoribbons, the width of the non-doted ribbons regions will decide the confinement effect for the electrons and holes. Generally, the smaller the width is, the stronger confinement effect we have and the larger mini-bandgap will be introduced to the graphene sample. Further, an increased mini-bandgap indicates a reduced bipolar effect that is causing a cancellation between the hole- and the electron-thermopower. Therefore, the mini-bandgap is negatively correlated with the interporous distance and positively correlated with the pore-radius. However, it is interesting to observe that the power factor, however, does not change significantly from sample to sample. This is suggesting that the cooling or the refrigeration performance of an antidoted graphene is not very sensitive to the distribution variance of either the pore-radius or the interporous distance. This trend is positive news for the quality control of products in large scale industrialization, which implies a potential for inexpensive manufacturing.

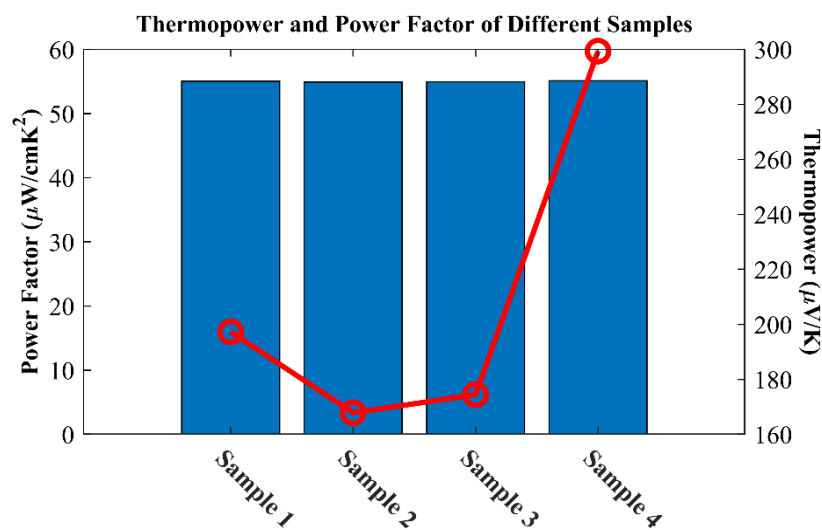

**FIGURE 3:** The thermopower of the four samples differ from each other because they have different pore-radii and/or interporous distances. However, the thermal power factors do not differ from sample to sample, noticeably, which implies easy quality control and inexpensive manufacturing for industrialization.

We now study the active rates of thermoelectric refrigeration and cooling, respectively. Since the four antidoted graphene samples have almost the same power factor at the optimized condition, we will just illustrate the results for Sample 1 as an instance. Figure 4 shows the data of the active rates for refrigeration and cooling using Sample 1 when the mid-point temperature is at the room temperature, namely, $T_{mid}=(T_H+T_C)/2=300$ K. The data for unoptimized cases and optimized cases are shown in Figure 4 (a) and (b), respectively. As we can see from the figure, the active rate for cooing is slightly higher than the refrigeration, but generally not noticeably different. The comparison between the optimized and unoptimized active rates are indicating an advantage of the low-dimensional materials tuned with gate voltage for microscale heat management. When thermoelectric refrigeration or cooling devices are designed, it is preferred that the *p*-leg and the *n*-leg have as similar performance as possible. However, as in the cases of most Peltier materials systems, the performance of *p*-type regime is different from the *n*-type region in the antidoted graphene samples as well. The hole-electron asymmetry ratio between the highest achievable active refrigeration/cooling rates for Sample 1 is ~1.17 at the room temperature. However, as in most carbon nanomaterials systems, this asymmetry ratio is generally smaller than the values in other commonly used large-*ZT* thermoelectric materials e.g. PbTe (8.79) [51] and BaSnO$_3$ (5.32×10$^3$) [52], which further implies a promising potential for the ultimate manufacturing of Peltier refrigerators and coolers based on the antidoted graphene samples.

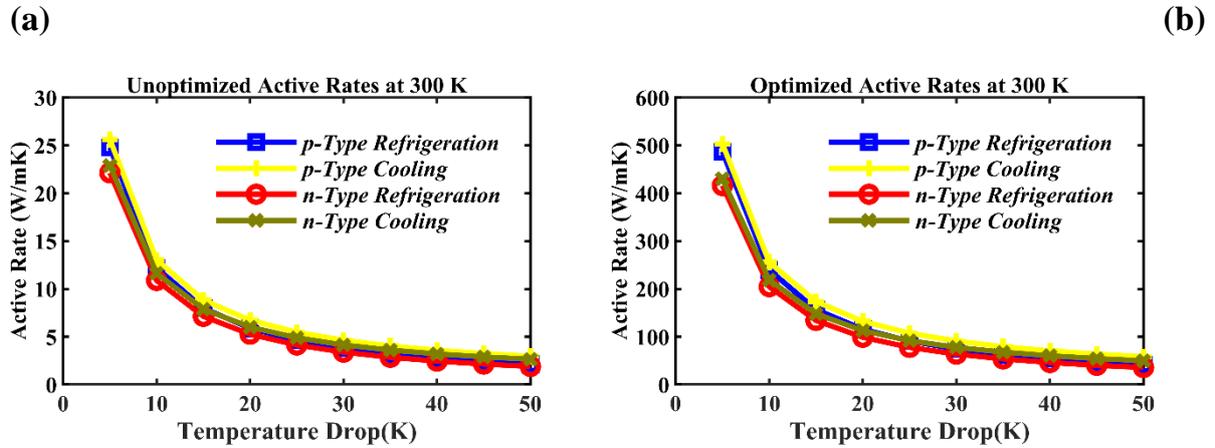

**FIGURE 4:** The (a) unoptimized and the (b) optimized active solid-state refrigeration and cooling out of the Peltier effect by Sample 1, for both the *p*-type and the *n*-type regimes. The mid-point temperature for all the data points is set at $T_{mid}$=300 K. The temperature drop marks the difference between the hot side and the cold side, i.e. $\Delta T=T_H-T_C$.

As in other carbon systems, the smaller the temperature drop is adopted, the higher performance is expected. When the temperature drop is 10 K, the active refrigeration and cooling rate can possible achieve as high as ~400 W/mk, which is similar to the thermal conductivity of copper [53]. When the temperature drop is widened, the active rate is generally declining. However, the refrigeration or cooling systems in practice are usually cascaded into different small-range temperature windows to ensure the overall performance.

Further, we have studied the active rates for Peltier refrigeration and cooling at a cryogenic temperature of 82 K. The data for unoptimized cases and optimized cases are shown in Figure 5 (a) and (b), respectively. The heat management at this temperature is generally more difficult, and the active rates are significantly smaller compared with the cases at the room temperature. However, the highest achievable refrigeration and cooling rates are can be ~300 times improved compared to the unoptimized value shown in Figure 5 (a). When we compare the cases of the four different samples, we can see that the performance is still not sensitive to the distribution variance of pore-radius and the interporous distance, which again is helpful for the industrialization and inexpensive manufacturing.

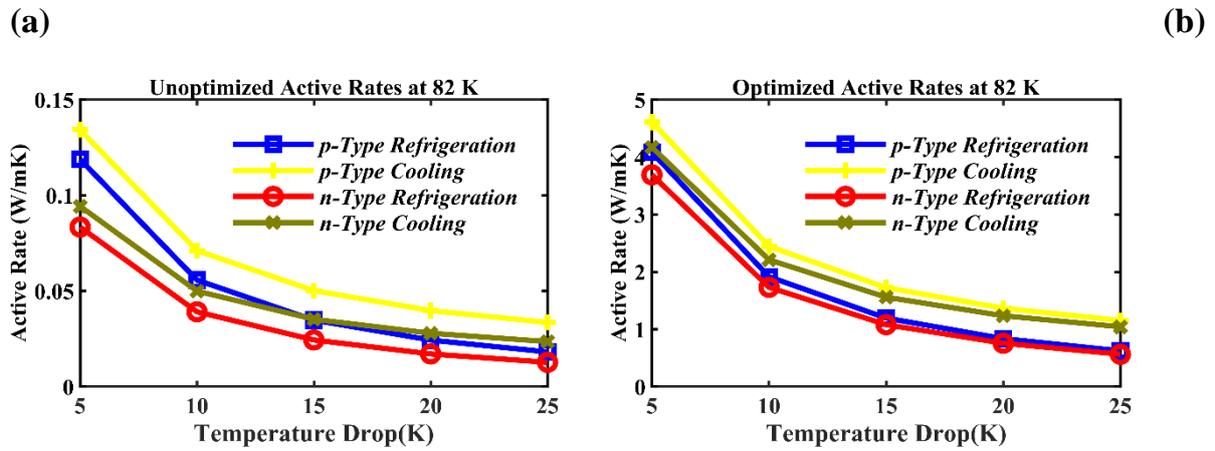

**FIGURE 5:** The (a) unoptimized and the (b) optimized active solid-state refrigeration and cooling out of the Peltier effect by Sample 1, for both the *p*-type and the *n*-type regimes. The mid-point temperature for all the data points is set at $T_{mid}$=82 K. The temperature drop marks the difference between the hot side and the cold side, i.e. $\Delta T = T_H - T_C$.

### 4. Conclusions

In conclusion, we have studied the optimized performance of antidoted graphene samples for the potential application in heat management of microelectronics. The active rate for refrigeration and cooling at the room temperature of 300 K and the cryogenic temperature of 82 K are investigated. We have found that, by considering the sensitivity of transport to carrier energy, the calculation of thermopower and thermal power factor can be more consistent with the experimental data, compared to the traditional methodologies including the *BoltzTrap* method and the *EPW* method. It is also interesting to see that though the pore-radius and the interpore distance can affect the thermopower, the optimal power factor and the active rates of cooling and refrigeration are, however, very robust against such variations that may be introduced during industrial scale manufacturing. The same methodology can also be used in other layered nanomaterials such as $MoS_2$, $WS_2$, $MoSe_2$, $WSe_2$, blue phosphorene, and tellurium.

**Acknowledge**

The authors acknowledge the Center for Computational Innovations at the Rensselaer Polytechnic Institute for providing the AIMOS supercomputer for our research and student training.

**Author Contributions**

S.T. designed and performed the research, analyzed the data, and wrote the paper. A. J., J.D.M., and J. G. helped in preparing, processing, and organizing the data.

**Conflict of interest**

There are no conflicts to declare.

**Table of Contents Entry:**

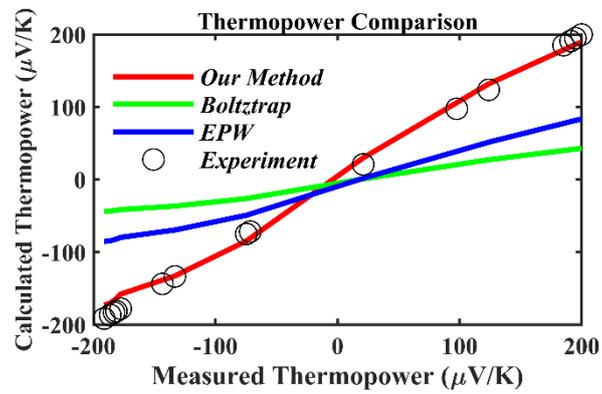

**20-word summary:**

Optimized heat management performance for microelectronics of antidoted graphene is studied using a new methodology with enhanced consistency with experiment.